\documentclass[prd,twocolumn,showpacs, showkeys,nofootinbib,floatfix]{revtex4-1}

\usepackage{amsmath}
\usepackage{amsfonts}
\usepackage{txfonts}
\usepackage{graphicx}
\usepackage{dcolumn}
\usepackage{bbm}
\usepackage{amssymb}
\usepackage{latexsym}
\usepackage{titlesec}
\usepackage[colorlinks=true, linkcolor=red, citecolor=blue]{hyperref}
\usepackage{longtable}

\begin{document}

\title{Detecting the Neutrinos Mass Hierarchy from Cosmological Data}

\author{Lixin Xu}
\email{lxxu@dlut.edu.cn}

\affiliation{Institute of Theoretical Physics, School of Physics \&
Optoelectronic Technology, Dalian University of Technology, Dalian,
116024, P. R. China}

\author{Qing-Guo Huang}
\email{huangqg@itp.ac.cn}

\affiliation{CAS Key Laboratory of Theoretical Physics, Institute of Theoretical Physics,
Chinese Academy of Sciences, Beijing 100190, China}

\affiliation{School of Physical Sciences, University of Chinese Academy of Sciences, No. 19A Yuquan Road, Beijing 100049, China}

\begin{abstract}

We propose a new parameterization to measure the neutrino mass hierarchy, namely $\Delta=(m_3-m_1)/(m_1+m_3)$ which is dimensionless and varies in the range $[-1,1]$. Taking into account the results of neutrino oscillation experiments, $\Delta$ is the unique parameter for determining all the masses of neutrinos, and a positive (negative) sign of $\Delta$ denotes the normal (inverted) mass hierarchy. Adopting the currently available cosmic observations, we find that the normal mass hierarchy is slightly favored, and the mass of lightest neutrino is less than $0.030$ eV for the normal mass hierarchy and $0.024$ eV for the inverted mass hierarchy at $95\%$ confidence level.     

\end{abstract}


\maketitle

\section{Introduction}
The neutrino oscillations imply that at least two neutrinos have non-zero masses \cite{ref:NOS1998}. However, up to now, we only measure the differences of neutrino mass squares in a standard scenario with three massive eigenstates \cite{ref:muMass}, i.e. 
\begin{eqnarray}
\Delta m^2_{21}\equiv m^2_2-m^2_1=7.5\times 10^{-5}\ \text{eV}^2,\label{eq:m212}\\
|\Delta m^2_{31}|\equiv |m^2_3-m^2_1|=2.5\times 10^{-3}\ \text{eV}^2. \label{eq:m312}
\end{eqnarray}   
The sign of $\Delta m^2_{31}$, i.e. the neutrino mass hierarchy, is still undetermined. Assuming that the mass of lightest neutrino is equal to zero, one can easily derive the lower bound on the total neutrino mass which are $0.06\ \text{eV}$  for the normal hierarchy (NH) and $0.10\ \text{eV}$ for the inverted hierarchy (IH). Comparing to the NH, the minimum total neutrino mass for IH is roughly doubled. NH can be distinguished from IH if the total neutrino mass is less than $0.1$ eV. Since gravity is much sensitive to the mass distribution, including the neutrino mass, the cosmological observations can play an important role to detect the neutrino mass hierarchy. Actually, massive neutrino cosmology had already been studied extensively in the past few years (see \cite{ref:NCbook} for a comprehensive review). 

In the early Universe long before the last scattering of the cosmic microwave background (CMB) photons, the neutrinos behave like cosmic radiations due to their small masses. The matter-radiation equality time will be altered by the different neutrino mass. This is the so-called early integrated Sacks-Wolfe (eISW) effect which affects the first peak of CMB temperature anisotropies and polarization power spectra. Subsequently massive neutrinos become non-relativistic and contribute to the total matter density in the Universe at both the background and perturbation levels during the matter and dark energy domination eras. Therefore the density fluctuations on small scales are washed out, while neutrinos behave as cold dark matter on large scales. In addition, the matter power spectrum is suppressed by a factor $\Delta P/P\approx -8f_{\nu}$ due to the lack of neutrino power, where $f_{\nu}$ is defined by $f_{\nu}\equiv\Omega_{\nu}h^2/\Omega_{m}h^2$ \cite{ref:HuNeutrino}. Therefore, the geometric and dynamic measurements are useful to determine the neutrino mass. 

Even though the neutrino mass has not been determined,  {\it Planck} 2015 already put a stringent  upper limit on the total neutrino mass, i.e. $\sum_\nu m_{\nu}<0.23\ \rm{eV}$ \cite{ref:planck2015CP}. Furthermore, future 21cm, precise CMB polarization observations \cite{ref:201621cm} and the cross correlation between the Rees-Sciama effect and weak lensing \cite{ref:Xuneutrino} are quite helpful to measure their masses. 

\begin{center}
\begin{figure}[tbh]
\includegraphics[width=8.0cm]{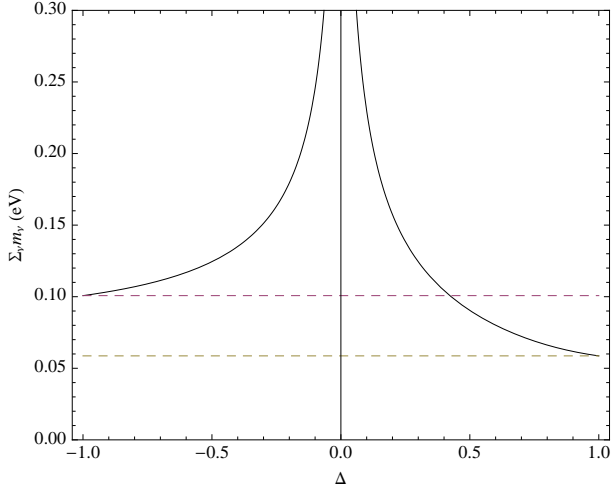}
\caption{The total neutrino mass.}\label{fig:summu}
\end{figure}
\end{center}

In this paper, we focus on the determination of neutrino mass hierarchy in cosmology. Recent study shows that the normal hierarchy is slightly favored by the cosmological observations \cite{ref:Huangneutrino} where the normal and inverted mass hierarchy are considered separately. Here we improve the ideas in \cite{ref:Huangneutrino} and try to treat both the normal and inverted mass hierarchy in a united way. We introduce a new parameter, named the neutrino mass hierarchy parameter, 
\begin{equation}
\Delta=\frac{m_3-m_1}{m_1+m_3} 
\end{equation}
to measure the neutrino mass hierarchy. The positive (negative) sign of $\Delta$ denotes the  normal (inverted) mass hierarchy. Similarly, see also in Ref. \cite{ref:Jimenez2010}. This mass hierarchy parameter is dimensionless and can vary in the range $[-1,1]$. Taking into account Eqs.~\eqref{eq:m212} and \eqref{eq:m312} from the neutrino oscillation experiments, the three neutrino mass eigenvalues are given by 
\begin{eqnarray}
m_1&=&\frac{1-\Delta}{2\sqrt{|\Delta|}}\sqrt{|\Delta m_{31}^2|},\\
m_2&=&\sqrt{m_1^2+\Delta m_{21}^2},\\
m_3&=&\frac{1+\Delta}{2\sqrt{|\Delta|}}\sqrt{|\Delta m_{31}^2|}.
\end{eqnarray}
In this sense we do not need to deal with the different neutrino mass hierarchies separately any more as that in \cite{ref:Huangneutrino}. 
The minimum eigenvalue of neutrino masses is $m_1$ for $\Delta>0$ (NH) or $m_3$ for $\Delta<0$ (IH). The lightest neutrino is massless if $\Delta=1$ for NH or $\Delta=-1$ for IH. 
From the above equations, the total neutrino mass is given by 
\begin{equation}
\sum_\nu m_\nu=\sqrt{\frac{\Delta m_{31}^2}{|\Delta|}}+\sqrt{\frac{(1-\Delta)^2}{4|\Delta|} |\Delta m_{31}^2|+\Delta m_{21}^2}\ . 
\end{equation} 
The total neutrino mass as a function of $\Delta$ is illustrated in Fig.~\ref{fig:summu}. 
We see that the total neutrino mass is not less than $0.06$ eV automatically. 
In the limit of $|\Delta|\rightarrow 0$, the total neutrino mass goes to infinity and cannot fit the data. 
Note that the dimensionless energy density for neutrinos is given by 
\begin{equation}
\Omega_{\nu}=\frac{\sum_\nu m_{\nu}}{93.14h^2\ \text{eV}},
\end{equation}
where $h$ is related to the Hubble constant by $H_0=100h \text{ km} \text{ s}^{-1}\text{ Mpc}^{-1}$. 

In Section \ref{sec:constraints}, we will use the currently available cosmological data to determine the neutrino mass hierarchy parameter $\Delta$, and show the constraints on the minimum neutrino mass and the total neutrino mass which are the derived model parameters from the mass hierarchy parameter $\Delta$. Conclusion is included in Section  \ref{sec:conclusion}.

\section{Hierarchy Parameter $\Delta$ Effects on CMB and Matter Power Spectra and the Fitting Results} \label{sec:constraints}

The effects on the CMB power spectra from massive neutrinos have been coded in {\bf CAMB} \cite{ref:CAMB} where the total neutrino mass is set to be not less than $0.06$ eV by hand. Adopting our new parametrization of neutrino mass in this paper, the total neutrino mass is not less than $0.06$ eV for NH and $0.1$ eV for IH automatically. 

\begin{center}
\begin{figure}[tbh]
\includegraphics[width=8.0cm]{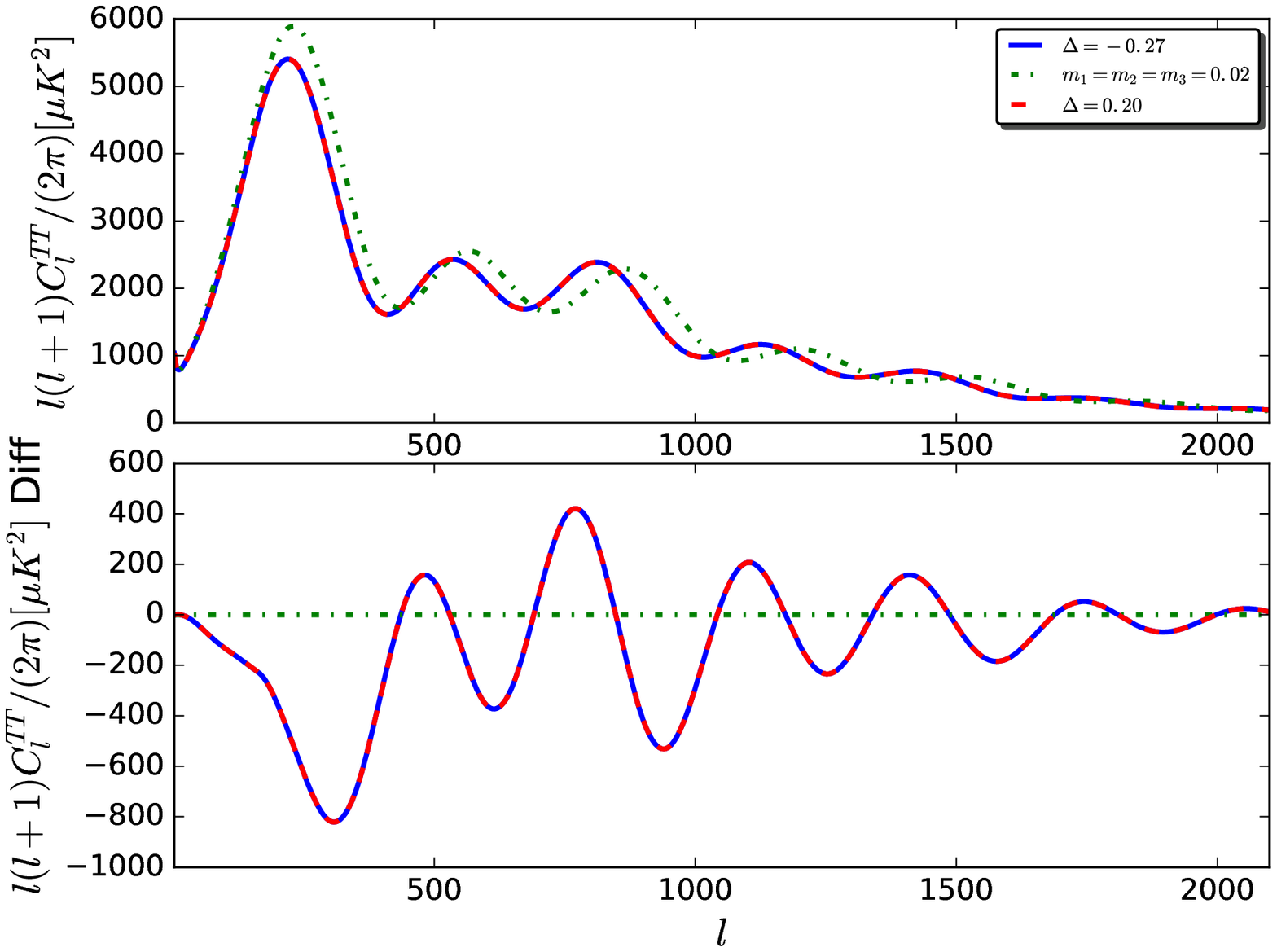}
\includegraphics[width=8.0cm]{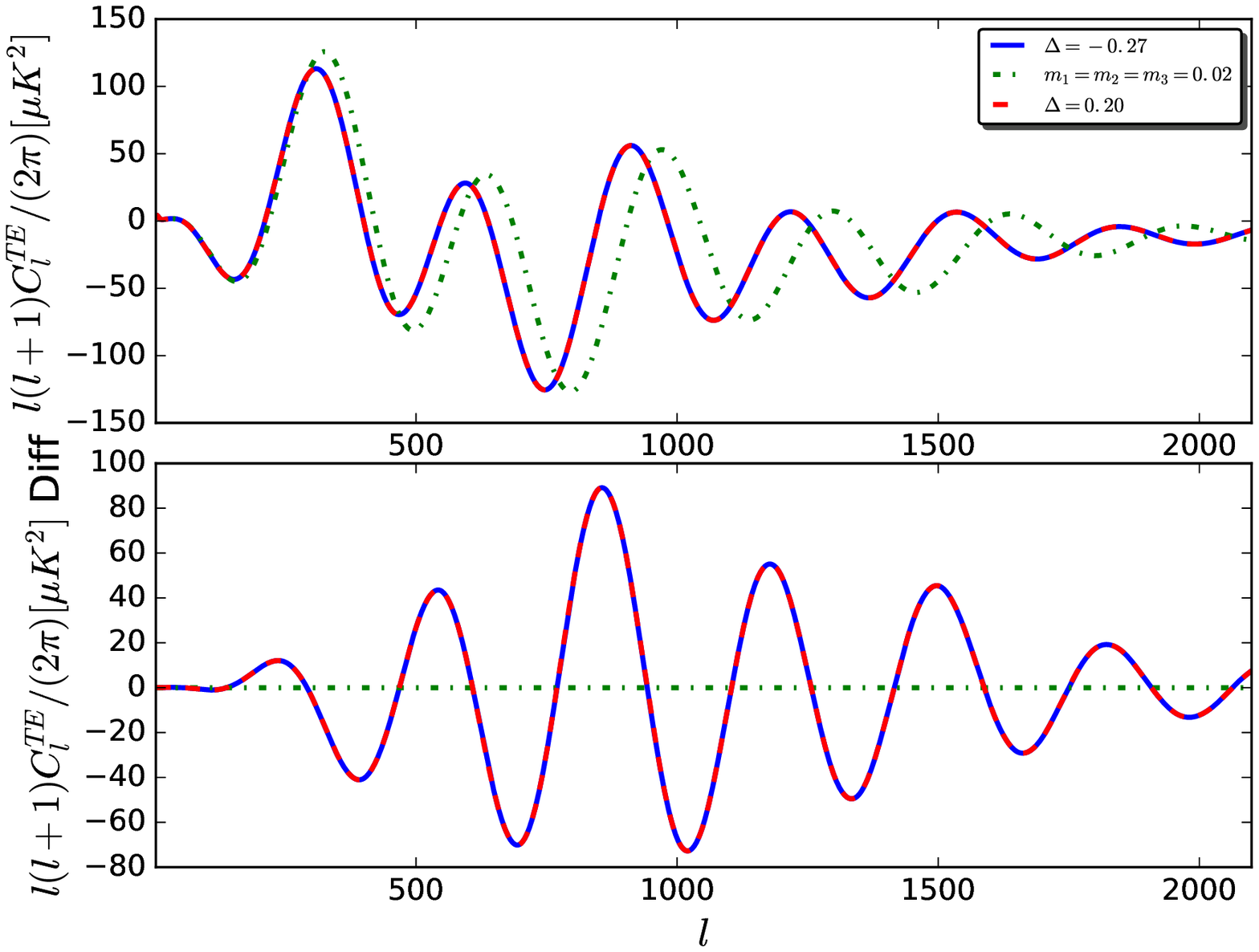}
\includegraphics[width=8.0cm]{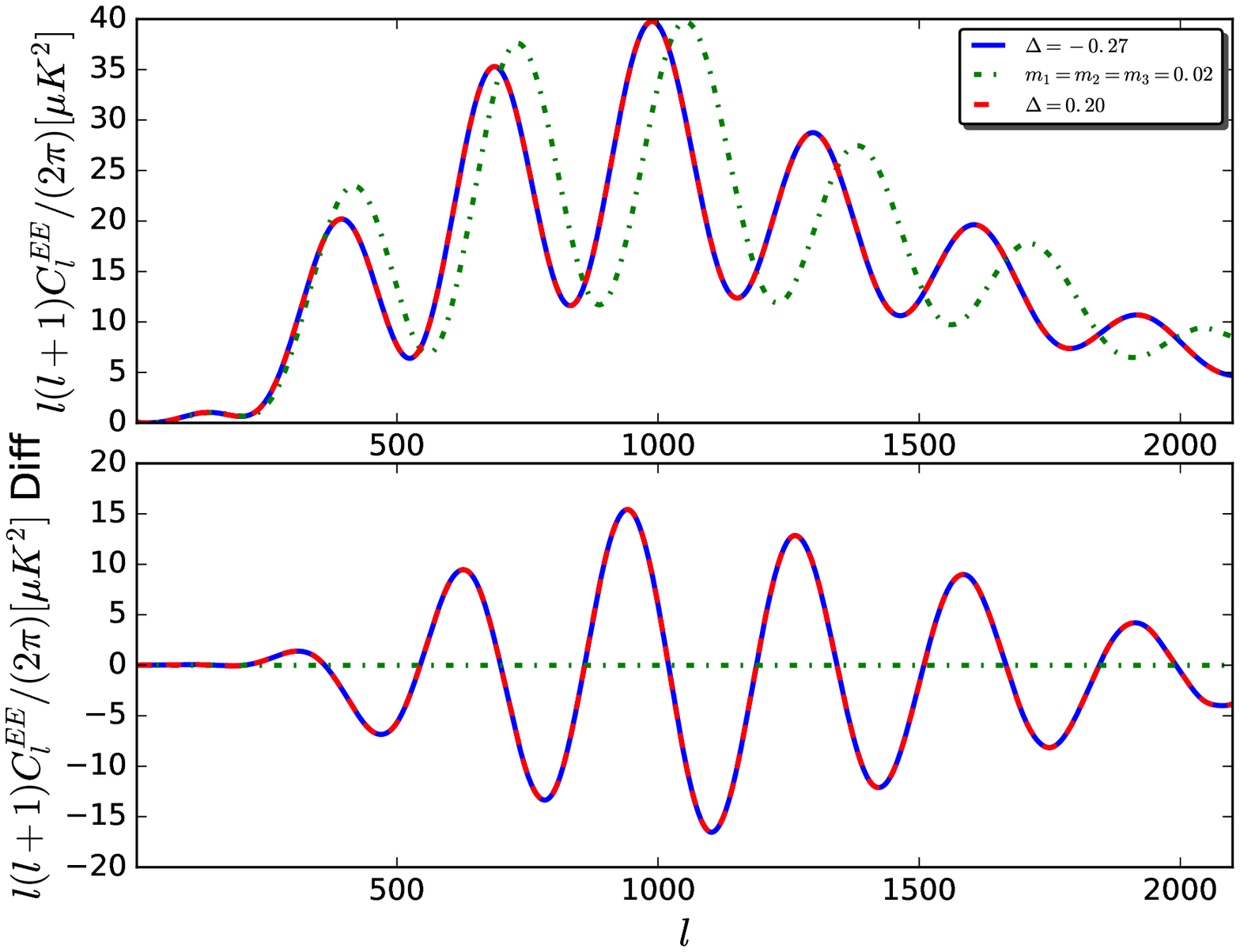}
\caption{The effects of hierarchy parameter $\Delta$ on the CMB TT, TE and EE power spectra from the top to the bottom respectively.}\label{fig:cmbeffects} 
\end{figure}
\end{center}

\begin{center}
\begin{figure}[tbh]
\includegraphics[width=8.0cm]{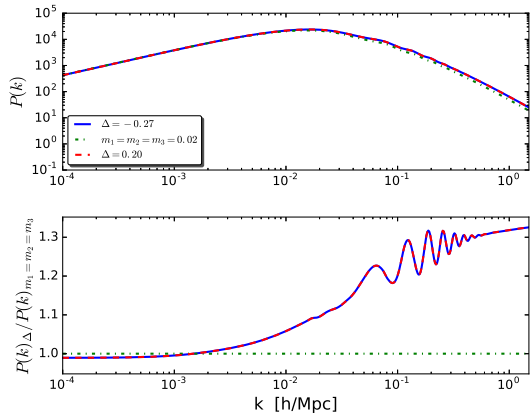}
\caption{The effects of hierarchy parameter $\Delta$ on the matter power spectrum at $z=0$.}\label{fig:pkeffects}
\end{figure}
\end{center}

In order to clarity the effects of neutrino mass hierarchy parameter $\Delta$ on the CMB and matter power spectra, we modified the publicly available {\bf CAMB} code to include the mass hierarchy parameter $\Delta$.
Adopting different values of $\Delta$, but keeping the other relevant cosmological  parameters fixed to their best fit values released by {\it Planck} 2015 \cite{ref:planck2015CP}, we show the effects on the CMB TT, TE and EE power spectra in Fig.~\ref{fig:cmbeffects} and the effects on the matter power spectrum at redshift $z=0$ in Fig.~\ref{fig:pkeffects}. 
Here the dash-dotted curves denote the power spectra for the cosmological model with $m_1=m_2=m_3=0.02$ eV. In order to compare the effects on the CMB and matter power spectra between NH and IH, we set $\Delta=-0.27$ for IH and $\Delta=0.2$ for NH respectively, and then the total neutrino masses are the same ($\sum_\nu m_\nu\simeq 0.157$ eV) for these two different mass hierarchies. From Figs.~\ref{fig:cmbeffects} and \ref{fig:pkeffects}, there are no any significant differences between NH and IH. It implies that both the CMB and matter power spectra are sensitive to the total neutrino mass, not the neutrino mass hierarchy. Furthermore, the neutrinos with larger total mass suppress the first peaks of CMB TT, TE, and EE power spectra and the matter power spectrum at large scales, but it enhances the matter power spectrum at small scales. 

In our cosmological model there are six based parameters and one extra neutrino mass hierarchy parameter, namely 
\begin{equation}
\{\omega_b,\omega_c, 100\theta_{MC},\tau, n_s, \ln (10^{10}A_s),\Delta\}, 
\end{equation}
where $\omega_b$ is the baryon density today, $\omega_c$ is the cold dark matter density today, $\theta_{MC}$ is the angular scale of the sound horizon at last-scattering, $\tau$ is the Thomson scattering optical depth, $A_s$ is the amplitude of scalar power spectrum, and $n_s$ is the spectral index of scalar power spectrum. 
Here the total neutrino mass $\sum_{\nu} m_{\nu}$ is a derived parameter. We use the data combination {\it Planck} 2015 LowTEB, TT, TE, EE + BAO DR12 + JLA SN + HST 2016 to constrain the cosmological parameters in our model. The {\it Planck} 2015 likelihood code is available online\footnote{https://wiki.cosmos.esa.int/planckpla2015/index.php/CMB\_spectrum\_\%26\_Likelihood\_Code} \cite{ref:planck2015likelihood}. The BAO data includes 6dFGS \cite{ref:6DFGS}, MGS \cite{ref:MGS} and BOSS DR12 CMASS and LOWZ \cite{ref:BAODR12}, and also the RSD data from CMASS and LOWZ \cite{ref:BAODR12}. For SN Ia, we use the Joint Light-curve Analysis (JLA) sample \cite{ref:SNJLA}. The recent value of the Hubble constant determined from SN Ia $H_0=73.03\pm1.79\rm{kms}^{-1}\rm{Mpc}^{-1}$ \cite{ref:HST2016} is also adopted, although there is some tension with {\it Planck} 2015 \cite{ref:planck2015CP} in the six-parameter based $\Lambda$CDM model.

\begin{center}
\begin{figure}[tbh]
\includegraphics[width=8.0cm]{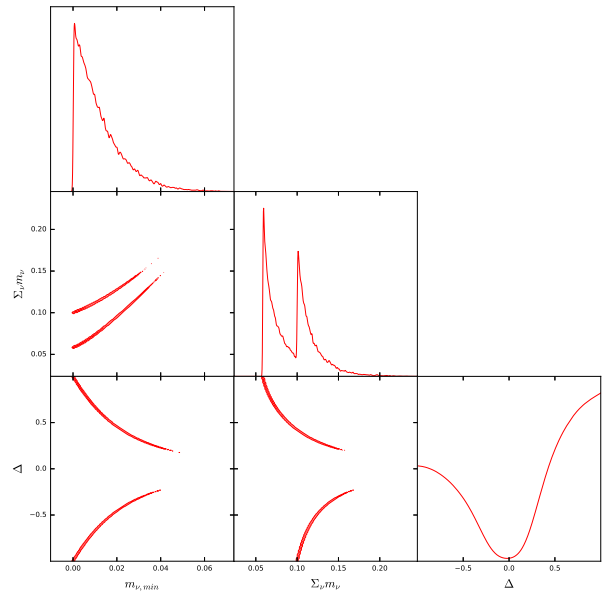}
\caption{The 1D marginalized distribution and 2D contours for interested model parameters with $68\%$ and $95\%$ confidence levels.}\label{fig:contour}
\end{figure}
\end{center}

We modify the {\bf CosmoMC} code \footnote{http://cosmologist.info/cosmomc} \cite{ref:cosmomc} to include the neutrino mass hierarchy parameter $\Delta$, and then run the program with eight independent chains. Our global fitting results are summarized in Tab.~\ref{tab:results} and Fig.~\ref{fig:contour}.
\begin{center}                                                                                                                  
\begin{table}                                                                                                                   
\begin{tabular}{ccc}                                                                                                            
\hline\hline                                                                                                                    
Parameters & Flat Priors & $68\%$ limits \\ \hline
$\Omega_b h^2$ & $[0.005,0.1]$ & $    0.02237\pm 0.00014$ \\
$\Omega_c h^2$ & $[0.001, 0.99]$ &$    0.1177\pm 0.0010$ \\
$100\theta_{MC}$ & $[0.5,10]$ & $    1.04095\pm 0.00031$ \\
$\tau$ & $[0.01,0.8]$ & $    0.078\pm 0.016$ \\
${\rm{ln}}(10^{10} A_s)$ & $[2,4]$ & $    3.086\pm 0.032$ \\
$n_s$ & $[0.8,1.2]$ & $    0.9697\pm 0.0038$ \\
\hline
$H_0$ & $[40,100]$ & $   67.93\pm 0.48$ \\
$\Omega_\Lambda$ & $-$ & $    0.6942\pm0.0062$ \\
$\Omega_m$ & $-$ & $    0.3058\pm 0.0062$ \\
$\sigma_8$ & $-$ & $    0.815\pm 0.013$ \\
$z_{\rm re}$ & $-$ & $    9.89_{-    1.36}^{+    1.57}$ \\
\hline 
$\Delta$ (95$\%$) & $[-1,1]$ &  $-1\leq \Delta<-0.40$ or $0.32<\Delta\leq 1$ \\
$m^{\rm{NH}}_{\nu,\rm{min}}$ eV (95$\%$) & $-$ & $    <0.030$ \\
$m^{\rm IH}_{\nu,\rm{min}}$ eV (95$\%$) & $-$ & $    <0.024$ \\
\hline\hline                                                                                                                    
\end{tabular}                                                                                                                   
\caption{The priors, $68\%$ limits and best fit values for the based and derived cosmological parameters.}\label{tab:results}                                                                                                   
\end{table}                                                                                                                     
\end{center}                                                                                                                    
The likelihood distribution for $\Delta$ in Fig.~\ref{fig:contour} shows that $\Delta>0$ (NH) is slightly favored compared to $\Delta<0$ (IH), and $\Delta=0$ is strongly disfavored. At $95\%$ confidence level (C.L.), $0.32<\Delta\leq 1$ for NH and $-1\leq \Delta<-0.40$ for IH. For NH, the mass of lightest neutrino is less than $0.030$ eV and the total neutrino mass is less than $0.119$ eV at $95\%$ C.L.. For IH, the mass of lightest neutrino is less than $0.024$ eV and the total neutrino mass is less than $0.135$ eV at $95\%$ C.L..

\section{Conclusion} \label{sec:conclusion} 
 
In this paper we propose a parameter ($\Delta=(m_3-m_1)/(m_1+m_3)$) to measure the neutrino mass hierarchy, namely the positive (negative) sign of $\Delta$ for normal (inverted) mass hierarchy. All of the neutrino masses are determined by $\Delta$ if the results of neutrino oscillation experiments are utilized. The two neutrino mass hierarchies can be treated in a united way according to this new parametrization. Unfortunately, we find that both the CMB and matter power spectra are sensitive to the total neutrino mass, not the mass hierarchy. However, the normal hierarchy is slightly preferred due to the fact that the lower total neutrino mass provides a slightly better fit to the current cosmological data. 

\acknowledgments

L.X is supported in part by National Natural Science Foundation of China under Grant No. 11275035, Grant No. 11675032 (People's Republic of China), and supported by ``the Fundamental Research Funds for the Central Universities" under Grant No. DUT16LK31. Q.-G.H is supported by Top-Notch Young Talents Program of China, grants from NSFC (grant NO. 11322545, 11335012 and 11575271), and Key Research Program of Frontier Sciences, CAS.

\end{document}